\documentclass[10pt,journal,compsoc]{IEEEtran}
\ifCLASSOPTIONcompsoc
  \usepackage[nocompress]{cite}
\else
  \usepackage{cite}
\fi
\usepackage{graphicx}
\ifCLASSINFOpdf
\else
\fi
\usepackage{amsmath,amsfonts}
\ifCLASSOPTIONcompsoc
  \usepackage[caption=false,font=footnotesize,labelfont=sf,textfont=sf]{subfig}
\else
  \usepackage[caption=false,font=footnotesize]{subfig}
\fi

\usepackage{stfloats}
\begin{document}
%
\title{AirCon: Over-the-Air Consensus for Wireless Blockchain Networks }
%
%
%
%

\author{Xin~Xie,
        Cunqing~Hua,~\IEEEmembership{Member,~IEEE,}
        Pengwenlong~Gu,~\IEEEmembership{Member,~IEEE,}
        \hfil\break and Wenchao~Xu,~\IEEEmembership{Member,~IEEE}
\IEEEcompsocitemizethanks{\IEEEcompsocthanksitem X. Xie and C. Hua are with the School of Cyber Science and Engineering, Shanghai Jiao Tong University, Shanghai 200240, China. \protect\\ E-mail: xiexin\_312@sjtu.edu.cn, cqhua@sjtu.edu.cn.
\IEEEcompsocthanksitem P. Gu is with the Department INFRES, LTCI, Telecom Paris, Institut Polytechnique de Paris, France. E-mail: gu@enst.fr.
\IEEEcompsocthanksitem W. Xu is with the Department of Computing, The Hong Kong Polytechnic University, Hong Kong, China. E-mail: wenchao.xu@polyu.edu.hk.}
}

\IEEEtitleabstractindextext{%
\begin{abstract}
Blockchain has been deemed as a promising solution for providing security and privacy protection in the next-generation wireless networks. Large-scale concurrent access for massive wireless devices to accomplish the consensus  procedure may consume prohibitive communication and computing resources, and thus may limit the application of blockchain in wireless conditions. As most existing consensus protocols are designed for wired networks, directly apply them for wireless users may exhaust their scarce spectrum and computing resources. In this paper, we propose AirCon, a byzantine fault-tolerant (BFT) consensus protocol for wireless users via the over-the-air computation. The novelty of AirCon is to take advantage of the intrinsic characteristic of the wireless channel and automatically achieve the consensus in the physical layer while receiving from the end users, which greatly reduces the communication and computational cost that would be caused by traditional consensus protocols. We implement the AirCon protocol integrated into an LTE system and provide solutions to the critical issues for over-the-air consensus implementation. Experimental results are provided to show the feasibility of the proposed protocol, and simulation results to show the performance of the AirCon protocol under different wireless conditions.
\end{abstract}

}

\maketitle

\IEEEdisplaynontitleabstractindextext

%
\IEEEpeerreviewmaketitle

\IEEEraisesectionheading{\section{Introduction}\label{sec:introduction}}

%
%
%
%
\IEEEPARstart{T}{he} next-generation wireless networks are expected to provide ubiquitous access to heterogeneous devices with ultra-high throughput, reliability, and extremely low latency \cite{YouXH_NSR_21}. This will bring great challenges to the security management in current 4G/5G mobile systems. Blockchain, a generic distributed ledger technology (DLT), has received extensive attention due to the significant advantages from  decentralization, immutability, and security in recent years. The blockchain technology enables registering and updating transactions in a decentralized fashion via consensus among participants, which has become the foundation of new security architecture for future wireless networks.

The consensus protocol plays an important role in the blockchain system. The efficiency of the consensus protocol determines blockchain system security bounds (fault tolerances) and performance such as transaction throughput, delay, and node scalability\cite{CaoB_Network_22}. Among many consensus protocols, the proof-based algorithm (PoX) is often used in a public chain, such as proof-of-work (PoW)\cite{Nakamoto_08}, proof-of-stake (PoS)\cite{PoS_Pavel}, whereby participants can join/leave the network without authentication. These PoX-based consensus protocols achieve good scalability with the cost of high resource consumption. For instance, the PoW protocol employed in Bitcoin consumes a huge amount of power resources to compute a meaningless hash value. And the confirmation delay and throughput of these consensus protocols also limit the application scenario. The voting-based consensus protocols,  such as PBFT\cite{Miguel_ACS_02}, RAFT\cite{Ongaro_ATC_14}, relies heavily on inter-participant communications to achieve  consensus. Therefore, the communication resource is critical for the voting-based protocols, which can only be applicable  for small/middle scale networks.

Most of the existing consensus protocols are primarily designed over wired communication networks, some new challenges may arise when they are deployed in wireless networks. First, wireless devices usually have limited resources (energy, storage, computation, etc.), so the resource-consuming PoX protocols are not suitable. Second, the wireless communication channel condition varies dynamically, not to mention that users often suffer from limited bandwidth resources, which significantly affects the consensus performance, especially for the voting-based protocols.

As a solution, the wireless channel features can be utilized to offer a solution to this dilemma. For the voting-based protocols, a block is generated based on the decision from the
majority participants\cite{Nguyen_JIPS_18}. For instance, in the PBFT protocol, each participant determines the consensus results by counting the number of messages with consistent hash from other participants. For traditional  ``communicate-then-consensus'' solutions, the consensus process can be conducted only after all messages from other participants are successful received and decoded. Instead of collecting messages from all participants, the consensus process only require the result of a function of these messages, e.g., the number of messages with consistent hash from other participants, rather than the details bits of each message. Therefore, a ``communicate-and-consensus'' scheme can be adopted to reduce the communication and computational complexity. Specifically, the communicating and consensus tasks can be accomplished simultaneously by leveraging the over-the-air computation (AirComp) technology\cite{Nazer_TIT_07} in wireless networks, which is a novel technology that exploits the superposition property of wireless multiple access channel (MAC) to aggregate signals from multiple participants over the same wireless channel. In this way, the destructive interferences from different participants are turned into a constructive one by appropriately matching the structure of the wireless channel and the function of the signals can be computed from the aggregated signal. 

In this paper, we propose ``AirCon'', an over-the-air consensus protocol for wireless networks, which significantly reduces the communication and computational overhead in a wireless network by using the AirComp technique. The novelty of this protocol is two-folded. First, the hash symbols of all participating users are encoded using lattice codes and transmitted to the base station (BS) simultaneously over the same wireless spectrum via the AirComp technique, which greatly reduces the wireless resource usage. Second, by leveraging the structural property of the lattice codes, each user can verify the consistency of its hash value with respect to the aggregated hash symbols without decoding the signal, whereby the consensus can be achieved with extreme low computational complexity. 

To show the feasibility of the AirCon protocol, we implement AirCon on an LTE system, where two key issues are solved: 1) Synchronization problem, which is the key to ensure that all participants transmit data in the same frequency simultaneously such that the signals can be accurately aligned at the BS. We will discuss the details of LTE synchronization mechanisms and show that these intrinsic mechanisms are sufficient for satisfying the synchronization requirement for the implementation of AirComp in an LTE system. 2) Uplink channel estimation and feedback problem. The participants need to learn the uplink channel state information (CSI) so that the channel fading can be pre-compensated. To this end, we propose a flexible reference symbol assignment scheme, which achieves a better tradeoff between estimation accuracy and estimation latency. 

The main contributions of the paper are summarized as follows:
\begin{itemize}
	\item We propose a novel AirComp-based consensus protocol (AirCon), which enables all participants to transmit their hash symbols to the BS simultaneously using the same wireless channel, and thus significantly reduces the communication cost of consensus messages. 
	\item We propose a hash consistency verification scheme in the physical layer, wherein each participant determines whether its hash value is consistent with the aggregated hash symbols or not without decoding the signal, which significantly reduces the complexity of consensus computation.	
	\item We implement AirCon based on srsLTE\footnote{The latest version of srsLTE has been renamed as srsRAN and to support 5G standards.}\cite{srsLTE}, an open-source LTE soft defined radio (SDR) platform. Extensive experiments are conducted to validate the practical performance of AirCon in a real-world LTE testbed. We also provide simulation results to evaluate the performance of the proposed schemes under more general network conditions.
\end{itemize}

The rest of this paper is organized as follows. In Section \ref{sec:related_works}, we discuss the research progress in blockchain-enabled wireless networks and the AirComp technique. In Section \ref{sec:system_model}, we introduce the system model and propose a modified PBFT-based over-the-air consensus (AirCon) procedure for wireless blockchain systems. The detailed design of AirCon protocol is presented in Section \ref{sec:aircon} and the details for implementation of AirCon in the LTE system are discussed in Section \ref{sec:aircomp}. Experimental and simulation results are provided in Section \ref{sec:performance}. 
Finally, we summarize this work in Section \ref{sec:conclusion}. 

\section{Related Works}\label{sec:related_works}
In this section, we give a brief introduction to the research progress on blockchain-enabled wireless networks and AirComp technique.
\subsection{Blockchain-enabled wireless networks} 
Blockchain has been studied extensively in literature as a security and privacy protection scheme for various application scenarios in wireless networks. In \cite{Christidis_Access_16}, the authors examined the application of blockchain in the Internet of things (IoT). They concluded that the blockchain-IoT combination is powerful and can lead to significant transformations across several industries. Blockchain technology also can be applied in the internet of vehicles\cite{Ortega_MVT_18} to provide cybersecurity protection for vehicular communications, including the dynamic control of source reliability, and the integrity and validity of the information exchanged. The incorporation of blockchain into the next-generation radio access network (RAN) also attracted great interest. A unified framework of the blockchain radio access network (B-RAN) was proposed in\cite{Ling_Access_19} as a trustworthy and secure paradigm for upcoming 6G networking.  Some critical elements of B-RAN, such as the deployment of smart contract\cite{Le_ICC_19}, trustworthy access\cite{Ling_Network_20,ZhangBW_WCSP_20}, mathematical modeling\cite{Ling_TCOMM_21} were also explored.

Due to the resource-limited feature of wireless devices, the blockchain system needs to be improved for the wireless environment. The authors in\cite{Ling_Access_19} proposed proof-of-devices (PoD) as a low-cost consensus protocol, which is based on the fact that B-RAN is comprised of a tremendous number of devices and attackers cannot control $51\%$ devices of the whole network. In an edge computing scenario, the authors in \cite{Guo_TVT_20} consider constructing a collaborative mining network (CMN) to execute mining tasks for mobile blockchain. Miners can offload their mining tasks to non-mining devices within a CMN when the resources are insufficient.  The authors in \cite{Liu_TII_19} proposed a lightweight blockchain system to reduce the computational cost and speed up the block generation rate in the industrial IoT (IIoT).

On the other hand, some research works were concerned with the impact of a wireless channel on blockchain performance. In\cite{Wei_Network_20}, the authors analyzed the trade-off between communication reliability and computing power in blockchain security and presented a lower bound to the computing power that is needed to conduct an attack with given communication reliability. Based on the widely used CSMA/CA mechanism, the impact of communication transmission delay on the confirmation delay, transaction per second, and transaction loss probability is analyzed in \cite{Cao_Network_20}. The authors in \cite{Sun_TIOTJ_19} analyzed the impact of node geographical distribution in the spatial domain and designed an algorithm to determine the optimal full-function node deployment. In \cite{CaoB_Network_22}, the authors evaluated the impact of scarce frequency spectrum resources on blockchain performance.

\subsection{Over-the-air computation} 

In this paper, we design a novel consensus protocol based on the AirComp technology, which exploits the intrinsical superposition property of the wireless channel to achieve efficient transmissions of multiple users.  AirComp was proposed by B. Nazer and M. Gastpar in their cornerstone work\cite{Nazer_TIT_07}, where an optimal joint source-channel strategy was developed to reliably reconstruct a function of
sources over a multiple-access channel. Early research works of AirComp took on the information-theoretic view and focuses on the achievable computation rate under structured coding schemes. The transmission strategy under different source distribution was explored in \cite{Gastpar_TIT_08,Wagner_TIT_08,Soundararajan_TIT_12}. The authors in \cite{Gastpar_TIT_08,Wagner_TIT_08} employed lattice codes to efficiently compute the sum of source signals over MACs. Some recent research efforts have been made to study AirComp from other aspects. The authors in \cite{Caoxw_TWC_20,Liuwc_TWC_20} considered a more general power-limited scenario where some devices cannot fully compensate for channel fading, then a threshold-based power-control policy was proposed to minimize the mean square error (MSE) of the whole system.  Multifunction computation was achieved by using AirComp in a multiple antenna system, beamforming strategies have been designed in \cite{Chen_TIoTJ_18,Zhu_TIoTJ_19} to achieve optimal performance. In\cite{Dongjl_TSP_20}, the blind AirComp technique was proposed for low-complexity and low-latency wireless networks, which does not require channel state information for AirComp.  

The AirComp technique can be applied in a wide range of application scenarios. In\cite{Sigg_ICIoT_12,Goldenbaum_SENSORS_14}, AirComp was proposed for fast data fusion, whereby the fusion center (FC) attempts to compute a specific function from the data of all sensor nodes (e.g., average reading). AirComp also has been extended for model aggregation in the distributed edge learning system \cite{Yangkai_TWC_19,Zhu_TWC_21,Mohammadi_TSP_20}, which is shown to reduce the communication latency without significant loss of the learning accuracy\cite{ZhuGX_WC_21}. In a multi-agent system, AirComp can be used for distributed consensus control\cite{Molinari_ECC_18,Molinari_TCNS_21} to reach an agreement over a set of variables of common interest, such as velocity, acceleration, and trajectory in vehicular platooning. Our work extends the application of AirComp to the consensus protocol in wireless blockchain systems.

Despite the extensive theoretical work in literature, the research on the  implementation issues for AirComp is relatively scarce. In\cite{Sigg_ICIoT_12} and \cite{Goldenbaum_SENSORS_14}, the authors proposed to modulate the value to be computed through AirComp as the mean value or power of a random sequence, so the accurate synchronization and CSI is not necessary for implementing the AirComp technique. In \cite{Omid_arxiv_16}, the authors proposed to achieve system  synchronization via the ``AirShare'' scheme \cite{Abari_INFOCOM_15}, and the uplink CSI was obtained by using channel reciprocity between uplink and downlink transmissions.  In \cite{Altun_TELFOR_17}, the uplink CSI was obtained by feedback from the fusion center. These implementation schemes were designed for analog function computation (AFC) in a wireless network. In this paper, we implement the AirComp technique based on the universal LTE system, which demonstrate that AirComp can be realized in the current digital communication systems.

\section{System model}\label{sec:system_model}
\begin{figure}
	\centering
	\includegraphics[scale=0.25]{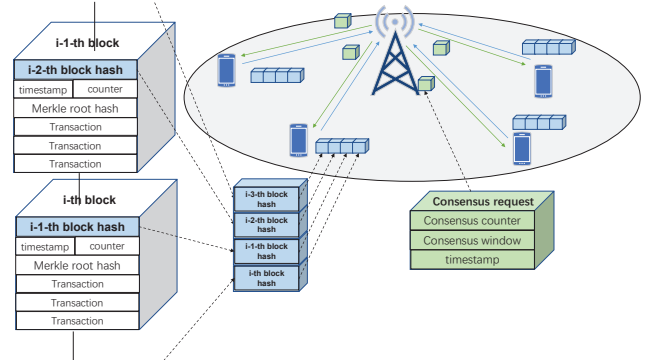}
	\caption{Blockchain-enabled wireless network model.}
	\label{fig_system_model}
\end{figure}

\begin{figure}
	\centering
	\includegraphics[scale=0.36]{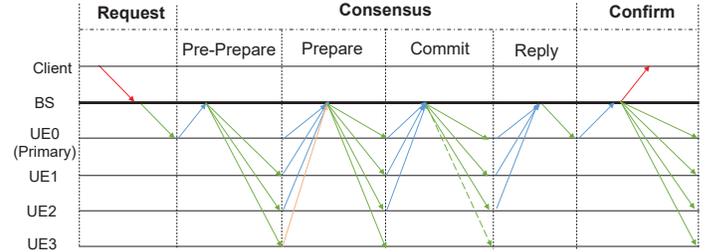}
	\caption{Diagram of the over-the-air consensus protocol. Hash verification for the generated block is achieved through the BS using the multiple access and broadcast channels.}
	\label{fig_aircon_pbft}
\end{figure}
We consider a blockchain-enabled wireless network as illustrated in Fig. \ref{fig_system_model}, where a set of $K$ users are served by a BS. These users maintain the same blockchain. The transaction blocks are generated periodically based on a consensus protocol, and stored in a distributed manner in all users.

In a blockchain system, the block appending procedure can be separated into three stages: Request, Consensus and Confirm, as shown in Fig. \ref{fig_aircon_pbft}. Our work focus on the consensus step and the other two steps are the same as the traditional blockchain systems.

{\bf Request}: the client sends a request to the primary user to trigger a transaction. After collecting enough transactions, the primary user generates a candidate block and broadcasts it to other replica users and requests to start the consensus procedure \footnote{We assume that the primary user is always honest. If other users or clients detect some malicious behavior from the primary user (such as broadcasting some illegal transactions), view change procedures\cite{Miguel_ACS_02} can be triggered to replace the primary user}.

{\bf Consensus}:  The consensus protocol in Fig. \ref{fig_aircon_pbft} is a revision of the PBFT protocol for wireless networks, which consists of four phases as follows:
\begin{itemize}
	\item Pre-prepare: After receiving the candidate block and request from the primary user, all replica users verify the transactions of the candidate block, then send the hash symbols of candidate blocks to the BS in the same uplink channel and enter the prepare phase. Note that user-specific information (such as user ID) should not be included in the generation of the hash value.
	
	\item Prepare and Commit: The BS obtains the superposed hash symbols and broadcasts them to all users. All replica users adopt a threshold-based two-round hash verification method to determine the consensus result.
	\item Reply: Only those users satisfying the threshold return the reply message to the primary user through the BS. If more than $\lceil{K/2}\rceil$ users return the message, the primary user determines that the consensus has been achieved.
\end{itemize}

{\bf Confirm}: If the consensus has been achieved, the primary user sends a confirm message to the BS, which is broadcast to all replica users and the client, then the candidate block is appended in each blockchain which is distributively stored at each replica user.

The bottleneck of the consensus protocol shown in Fig. \ref{fig_aircon_pbft} lies in the \emph{Prepare} and the \emph{Commit} phases. In the traditional hash cross-validation, the hash value generated by different users should be transmitted to each other via orthogonal wireless channels (such as TDMA, FDMA, or OFDMA in 4G/5G networks). Therefore, the wireless resource usage in these two phases increases with the number of participating users, which cannot guarantee reliable and low latency consensus for larger network size.

\section{AirCon Protocol Design}\label{sec:aircon}

In this section, we propose a novel over-the-air consensus (AirCon) protocol to address the transmission bottleneck problem in the \emph{Prepare} and the \emph{Commit} phases based on the AirComp and the lattice coding techniques. The idea is to modulate the hash bits of different users using the lattice codes, which can be transmitted simultaneously to the BS over the same wireless channel. Due to the structural property of the lattice codes, the BS can adopt the AirComp technique to aggregate the linear combination of all codes (which is a codeword itself), and forwards it back to all users, which can be used by each user to make a consensus decision without decoding the signal. 

In the following, we first introduce the preliminary background on the AirComp and lattice coding techniques. We then present the hash consistency verification scheme based on these two techniques, with which a two-round consensus procedure for the AirCon protocol is designed. 

\subsection{Over-the-air computation}
We consider an OFDM-based wireless system, which consists of a single BS and a set of $K$ users. All users transmit signals to the BS through the same uplink channel. The received signal at the BS is expressed as:
\begin{equation}\label{eq1}
	\mathbf{y}=\sum_{k=0}^{K-1}\mathbf{H}_k\mathbf{B}_k\mathbf{x}_k,
\end{equation}
where $\mathbf{x}_k\in\mathbb{C}^{N\times 1}$ is the symbol vector transmitted by user $k$ with the covariance matrix $\mathbb{E}\{\mathbf{x}_k\mathbf{x}_k^H\}=\mathbf{I}$, where $\mathbf{I}$ is the identity matrix.  $\mathbf{B}_k=\texttt{Diag}\{b_{k0},b_{k1},\cdots,b_{k N-1}\}\in\mathbb{C}^{N\times N}$ is the pre-processing matrix for user $k$. $\mathbf{H}_k=\texttt{Diag}\{h_{k0},h_{k1},\cdots,h_{k N-1}\}\in\mathbb{C}^{N\times N}$ is the channel matrix from user $k$ to the BS.  $N$ is the total number of sub-carriers.  $K$ is the total number of users. The superscript $(\cdot)^H$ denotes Hermitian transpose. 

Under the condition that each user $k$ has perfect knowledge of $\mathbf{H}_k$, and its transmit power is not bounded, then the channel fading can be pre-compensated by a simple channel-inversion operation as $\mathbf{B}_k=\mathbf{H}_k^{-1}$. In this way, the signals transmitted by all users will be aggregated at the BS as follows:
\begin{equation}\label{eq2}
	\mathbf{y}= \sum_{k=0}^{K-1}\mathbf{x}_k, 
\end{equation}
which is equivalent to the computation of the summation of all $\mathbf{x}_k$s over the wireless channel.  

Motivated by this property, the AirComp technique can be generalized to support a wide range of mathematical operations, which are based on the property of nomographic function in the following form:
\begin{equation}\label{eq3}
	f(\mathbf{s}_0,\mathbf{s}_1,...,\mathbf{s}_{K-1})=\psi{\bigg(}\sum_{k=0}^{K-1}\phi{(\mathbf{s}_k)}{\bigg)},
\end{equation}
where $\phi(\cdot)$ and $\psi(\cdot)$ denote pre- and post-processing functions, respectively. 

Some well-known nomographic functions include:
\begin{enumerate}
	\item Arithmetic Mean: $f(s_0,s_1,...,s_{K-1})=\frac{1}{K}\sum_{k=0}^{K-1}s_k$, with $\phi(s_k)=s_k$ and $\psi(y)=y/N$;
	\item Euclidean Norm: $f(s_0,s_1,...,s_{K-1})=\sqrt{\sum_{k=0}^{K-1}s_k^2}$, with $\phi(s_k) = s_k^2$ and $\psi(y)=\sqrt{y}$; 
	\item Number of Active Node: $f(s_0,s_1,...,s_{K-1})$ is the number of active node, with $\phi(s_k) = 1\,(\text{active})\,\text{or}\, 0\,(\text{inactive)}$ and $\psi(y)=y$.
\end{enumerate}

\subsection{Nested lattice code}
A lattice is an infinite discrete set of points in the Euclidean space that are regularly arranged and are closed under addition\cite{Natarajan_TIT_18}. As an important channel coding technique, the structural properties of nested lattice coding are well suited for multiple access channel in wireless networks, which allows multiple transmitters to effectively share the same radio resources and can protect against channel noise. Specifically, a $d$-dimensional lattice $\Lambda$ in the Euclidean space $\mathbb{R}^d$ can be generated as follows:
\begin{equation}\label{eq4}
	\Lambda=\{\mathbf{Gu}\ :\ \textbf{u}\in\mathbb{Z}^d\},
\end{equation}
where $\mathbf{G}=[\mathbf{g}_1,\mathbf{g}_2,...,\mathbf{g}_d]$ is a full-rank generator matrix.

A lattice $\Lambda_C$ is \emph{nested} in some lattice $\Lambda_F$ if $\Lambda_C\subseteq  \Lambda_F$, i.e., $\Lambda_C$ is a sublattice of $\Lambda_F$. In this case, $\Lambda_F$ is denoted as the fine lattice, which defines the codewords, while $\Lambda_C$ is denoted as the coarse lattice, which is used for shaping. Specifically, the \emph{nested lattice code} $\mathcal{L}$ is the set of all points of a fine lattice $\Lambda_F$ that is within the fundamental Voronoi region $\mathcal{V}_C$ of a coarse lattice $\Lambda_C$:
\begin{equation}\label{eq5}
	\mathcal{L}=\Lambda_F\cap\mathcal{V}_C=\{\textbf{x}\ :\ \textbf{x}=\lambda\ \text{mod}\ \Lambda_C,\lambda\in\Lambda_F\},
\end{equation}
where the \emph{fundamental Voronoi region}, $\mathcal{V}_C$, of the lattice $\Lambda_C$, is the set of all points in $\mathbb{R}^d$ that are closest to the zero vector:
\begin{equation}\label{eq6}
	\mathcal{V}_C=\{\textbf{z}\ :\ ||\textbf{z}||\le||\textbf{z}-\lambda||,\forall\lambda\in\Lambda_C,\textbf{z}\in\mathbb{R}^d\},
\end{equation}

For each user $k$, a $d$-dimension nested lattice codeword ${\bf{x}}_k\in\mathcal{L}$ can be generated based on its hash value ${\bf{s}}_k\in\mathbb{F}^l_p$ by the encoding function $\phi(\cdot)$ as follows:
\begin{equation}\label{eq7}
	\begin{aligned}
		\phi({\bf{s}}_k): \mathbb{F}^l_p&\rightarrow\mathbb{R}^d\\
		{\bf{s}}_k&\rightarrow{\bf{x}}_k
	\end{aligned}
\end{equation}
where $\mathbb{F}_p=\{0,1,...,p-1\}$ forms a finite field under integer arithmetic modulo $p$. Then for a nested lattice codebook $\mathcal{L}$, the following property is held for all ${\bf{x}}_k\in\mathcal{L}$:
\begin{equation}\label{eq8}
	{\bigg [}\sum_{k=0}^{K-1}{\bf{x}}_k\mod{\Lambda}_C{\bigg ]}\in\mathcal{L},
\end{equation}
that is,  the sum of lattice codewords modulo the shaping lattice is a codeword itself. Due to this linearity preserving characteristic, there exists a post-processing function $\psi(\cdot)$ that satisfies\cite{Nazer_TIT_11}:
\begin{equation}\label{eq9}
	\psi{\bigg (}\sum_{k=0}^{K-1}\phi({\bf{s}}_k)\mod{\Lambda_C}{\bigg )}=\underset{k=0}{\overset{K-1}{\bigoplus}}{\bf{s}}_k,
\end{equation}
which is  a nomographic function, that is,  $f({\bf{s}}_0,{\bf{s}}_1,...,{\bf{s}}_{K-1})=\underset{k=0}{\overset{K-1}{\bigoplus}}{\bf{s}}_k$.  

Based on this property, AirComp can be used to transmit and compute the superposition of hash values from all users. Specifically, each user $k$ maps its hash value $\mathbf{s}_k$ to a lattice codeword $\mathbf{x}_k$ using the mapping operation in Eq.\eqref{eq7}, then all users can transmit their lattice codes simultaneously to the BS using the AirComp technique. 
At the BS, the received signal is $\mathbf{y}=\sum_{k=0}^{K-1}\mathbf{x}_k+\mathbf{w}$, which is a $d$-dimension linear combination of the lattice codes of all users with noise of $\mathbf{w}$.

In theory, there are many choices for the lattice codes with different codeword dimensions. However, in practice, the transmitted symbol can only represent information in a 2-dimensional space (in-phase and quadrature dimension). Therefore, for the high dimensional hash value, such as the 128-bits hash generated by the MD5 algorithm\cite{Rivest_92}, it needs $2^{64}$ codewords at each dimension to represent the hash, which is impossible in a real communication system. 
\begin{figure}
	\centering
	\includegraphics[scale=0.5]{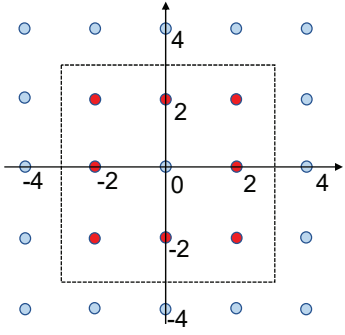}
	\caption{Nested  $\mathbb{Z}^2$ cubic lattice.}
	\label{fig_cubic_lattice}
\end{figure}

In this work, we consider an $\mathbb{Z}^2$ cubic lattice  $\tilde{\Lambda}_{F}$ as illustrated in Fig. \ref{fig_cubic_lattice},
{where the points are $\tilde{\Lambda}_{F}$ points and the Voronoi region of the coarse lattice $\tilde{\Lambda}_{C}$ is drawn in dashed lines.} 
Only the eight codes in the outer tier of the Voronoi region are used (shown in red color), so the code rate is $B=\log_2 8=3$ bits/symbol. Therefore, for an $L$-bits hash value, it can be partitioned into $N=\lceil{L/3}\rceil$ symbols.

\subsection{Hash consistency verification}\label{con_protocol}
Based on the $\mathbb{Z}^2$ cubic lattice in Fig. \ref{fig_cubic_lattice}, each user $k$ can construct a hash symbol vector $\tilde{\bf{x}}_k=[\tilde{x}_{k0},\tilde{x}_{k1},...,\tilde{x}_{k N-1}]^T$, where $\tilde{x}_{kn}$ is mapped to one of the eight codes. Then each element of $\tilde{\bf{x}}_k$ is transmitted in an OFDM sub-carrier. The received symbol at the BS in sub-carrier $n$ is $\tilde{y}_n=\sum_{k=0}^{K-1}\tilde{x}_{kn}+w_n$. 

The received symbol at sub-carrier $n$ can be quantized to the nearest point $t_n\in\mathbb{Z}^2$ in the $\tilde{\Lambda}_{F}$ as follows:
\begin{equation}\label{eq10}
	{t_n}=\arg\underset{\lambda\in{\tilde{\Lambda}_{F}}}{\min}||\tilde{y}_n-\lambda||,\,n\in[0,N-1], 
\end{equation}
which are then broadcasted to all users through the downlink channel. A vector $\mathbf{t}=[t_0,t_1,...,t_{N-1}]^T\in\mathbb{Z}^{2N}$ is constructed by each user, which is the complete linear combination of the lattice codes of all users. 

Based on the received vector $\mathbf{t}$, we design a consensus algorithm by exploiting the geometric characteristics of lattice codes. Let $C_m$ denote a set of $m$ users that transmit the consistent hash symbol vector, and $C_{K-m}$ denote the rest $K-m$ users that transmit different hash symbol vectors. Then $\mathbf{t}$ can be re-written as follows:
\begin{equation}\label{eq11}
	\mathbf{t}=\sum_{k\in C_m}\tilde{\mathbf{x}}_k+\sum_{k\in C_{K-m}}\tilde{\mathbf{x}}_k+\mathbf{e},
\end{equation}
where $\mathbf{e}$ is the random error pattern after symbols quantization in \eqref{eq10} with $\mathbb{E}\{\mathbf{e}\}=\mathbf{0}$. 

We assume that the hash symbol vector elements are uniformly mapped from the lattice codewords as shown in Fig. \ref{fig_cubic_lattice}. For each user $k\in C_m$, we assume the transmitted hash symbol vector is $\bar{\mathbf{x}}$, i.e., $\tilde{\mathbf{x}}_k=\bar{\mathbf{x}}, \forall k\in C_m$. For each user $k\in C_{K-m}$, we firstly consider the case that the hash symbol vector is independent of each other. Then for $k\neq j$, we have
\begin{equation}\label{eq14}
	\mathbb{E}\{\tilde{\mathbf{x}}_k^T\tilde{\mathbf{x}}_j\}=
	\begin{cases}
		N\sigma_s^2,&{\text{If}~\forall k,j\in C_m}\\
		0,&{\text{Otherwise}}
	\end{cases}
\end{equation}
where $\sigma_s^2$ is the variance of codewords. Furthermore, we define $\tilde{I}_{k}$ as the hash consistency factor (HCF) of user $k$, which is the normalized inner product of its hash symbol vector $\tilde{\mathbf{x}}_k$ and vector $\mathbf{t}$, that is:
\begin{equation}\label{eq13}
	\begin{aligned}
		\tilde{I}_k=\frac{\mathbf{t}^T\tilde{\mathbf{x}}_k}{K|\tilde{\mathbf{x}}_k|^2}, 
	\end{aligned}
\end{equation}

From \eqref{eq14}, we can obtain the expectation of $\tilde{I}_k$ as follows: 
\begin{equation}\label{eq20}
	\mathbb{E}\{\tilde{I}_k\}=
	\begin{cases}
		\frac{m}{K},&{\forall k\in C_{m}}\\
		\frac{1}{K}.&{\forall k\in C_{K-m}}
	\end{cases}
\end{equation}

\begin{figure}
	\centering
	\includegraphics[scale=0.22]{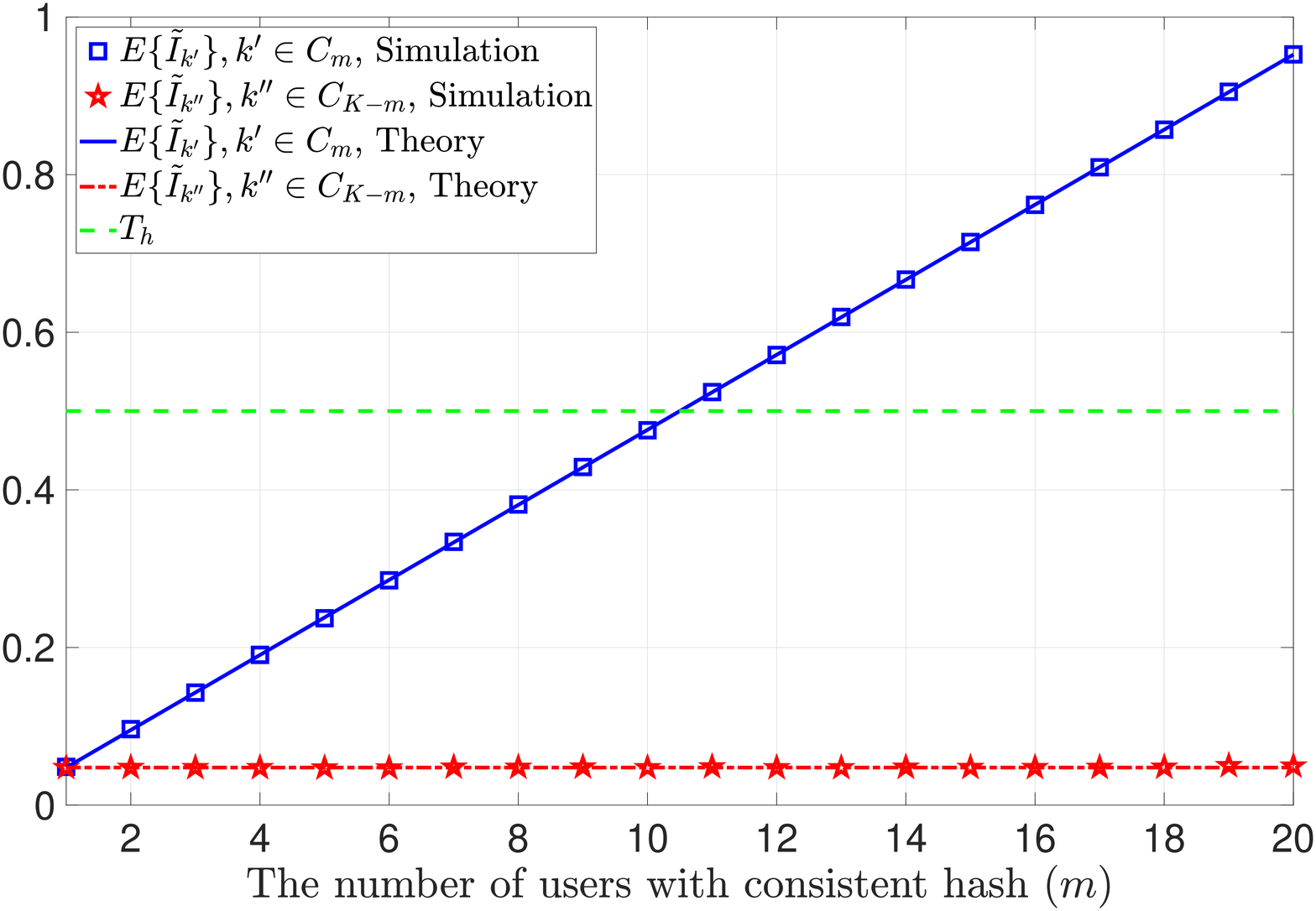}
	\caption{ $\mathbb{E}\{\tilde{I}_{k'}\}$ and $\mathbb{E}\{\tilde{I}_{k''}\}$  vs. $m$ ($K=21$).}
	\label{fig_threshold}
\end{figure}

In Fig.~\ref{fig_threshold}, we plot $\mathbb{E}\{\tilde{I}_{k}\}$ with 21 users ($K=21$) for $m$ varying from 1 to 20. It can be seen that for $m\geq 2$, $\mathbb{E}\{\tilde{I}_{k'}\}>\mathbb{E}\{\tilde{I}_{k''}\},\forall k'\in C_{m}, k''\in C_{K-m}$. Thus $\tilde{I}_{k}$ can be used as a reliable metric for each user to determine whether its hash symbol vector is consistent with the majority of hash vectors in $\mathbf{t}$ or not, that is, it belongs to one of the $m$ users sending the same hash or not. Therefore, this metric can be used to check if the consensus is reached or not. Specifically, in the proposed AirCon protocol, the consensus can be reached if $m>\lfloor{K/2}\rfloor$ users generate the same hash vector. According to \eqref{eq20}, we can set a threshold as $T_h = 0.5$. After receiving the superposed hash vector $\mathbf{t}$, each user $k$ calculates $\tilde{I}_k$ as defined in \eqref{eq13}. If $\tilde{I}_k\geq T_h$, the user assumes that the consensus can be achieved and sends the result to the primary user. 

However, if some users are controlled by an attacker, the consensus performance will be degraded based on this ideal threshold. For instance, consider the case where $m$ users are honest and the rest of $K-m$ users are controlled by a malicious attacker. The hash symbol vector transmitted by honest users is $\bar{\mathbf{x}}$, while the hash symbol vector transmitted by malicious users is $\hat{\mathbf{x}}=-\bar{\mathbf{x}}$, then the aggregated vector at the BS is $\mathbf{t}=m\bar{\mathbf{x}}+(K-m)\hat{\mathbf{x}}=(2m-K)\bar{\mathbf{x}}$. In this case, $\tilde{I}_k=(2m-K)/K$ for a honest user $k$, so the consensus cannot be achieved with a threshold of $T_h=0.5$ unless $m> \lfloor{3K/4}\rfloor$.  Therefore, this one-round consensus procedure is vulnerable to this kind of conspiracy attack. 

As a solution, we propose a two-round consensus procedure to improve the security of the proposed consensus procedure. Specifically, in the first round, malicious users can be filtered by setting a proper threshold and only honest users enter the second round to decide the consensus results. In the following, we will analyze the optimal threshold for the first round and the corresponding fault tolerance of the proposed consensus protocol.

Before getting into the details, we first make some assumptions about the capabilities of an attacker: 
\begin{enumerate}
	\item \emph{AS}$_1$: An attacker knows the hash symbols transmitted by each honest user. This is feasible because the attacker can calculate the legal hash by acting as an honest user;
	\item \emph{AS}$_2$: An attacker cannot manipulate the parameters of the physical layer, such as the transmitting power.  
	\item \emph{AS}$_3$: An attacker cannot manipulate the consensus protocol, such as the consensus threshold.
\end{enumerate}

We assume that an attack is successful if the consensus cannot be achieved even if the number of honest users is more than half of the total number of users (i.e., $m>\lfloor{K/2}\rfloor$).

Based on the aforementioned attacking model, we assume all honest users transmit $\bar{\mathbf{x}}$ and all malicious users transmit $\hat{\mathbf{x}}$. Then $\mathbf{t}$ can be re-written as follows:

\begin{equation}\label{eq16}
	\mathbf{t}=m\bar{\mathbf{x}}+(K-m)\hat{\mathbf{x}}+\mathbf{e},
\end{equation}
and the expectation of $\tilde{I}_k$ can be re-written by
\begin{equation}\label{eq18}
	\mathbb{E}\{\tilde{I}_k\}=
	\begin{cases}
		1-\alpha+\alpha\rho,&{\forall k\in C_{m}}\\
		\alpha+(1-\alpha)\rho,&{\forall k\in C_{K-m}}
	\end{cases}
\end{equation}
where $\alpha = \frac{K-m}{K}$ is the percentage of malicious users and  $\rho=\mathbb{E}\{\frac{\hat{\mathbf{x}}^T\bar{\mathbf{x}}}{|\bar{\mathbf{x}}|^2}\}=\mathbb{E}\{\frac{\bar{\mathbf{x}}^T\hat{\mathbf{x}}}{|\hat{\mathbf{x}}|^2}\}$ denote the correlation coefficient between $\hat{\mathbf{x}}$ and $\bar{\mathbf{x}}$. The attacker can change $\tilde{I}_k$ of the honest users by manipulating $\rho$ between $[-1, 1]$ with properly setting of  $\hat{\mathbf{x}}$. 

In the first round of consensus, a threshold should be set such that all honest users can get into the second round while the malicious users should be filtered as many as possible. To this end, the threshold should be set to
\begin{equation}\label{eq19}
	T_{h1}=\underset{\rho\in[-1,1]}{\text{min}}(1-\alpha+\alpha\rho)=1-2\alpha
\end{equation}

From the attacker's perspective, it must find a $\rho$ satisfying $\alpha+(1-\alpha)\rho>T_{h1}=1-2\alpha$ such that malicious users also can get into the second round. Otherwise, all malicious users will be filtered in the first round. Therefore, the range of $\rho$ can be given by
\begin{equation}\label{eq21}
	\rho>\frac{1-3\alpha}{1-\alpha}
\end{equation}

In the second round of consensus, the HCF $\tilde{I}_k$ of honest users should be more than $T_{h2}=0.5$ such that the consensus can be achieved even if all malicious users satisfy the threshold in the first round of consensus by setting $\rho$ as in \eqref{eq21}, that is
\begin{equation}\label{eq22}
	\tilde{I}_k=1-\alpha+\alpha\rho=1-\alpha+\alpha\frac{1-3\alpha}{1-\alpha}>\frac{1}{2},~\forall k\in C_m
\end{equation}

Then we can obtain the range of $\alpha$ as follows:
\begin{equation}\label{eq23}
	\alpha<\frac{\sqrt{17}-1}{8}\approx0.39:=\alpha^*
\end{equation}	

Substituting \eqref{eq23} into \eqref{eq19},  we can obtain the optimal value of $T_{h1}$ as follows:
\begin{equation}\label{eq24}
	T_{h1}^*:=0.22
\end{equation}	

The range of $\alpha$ in \eqref{eq23} specifies the fault tolerance of the two-round consensus protocol. If the percentage of malicious users does not exceed 0.39, then the consensus can be achieved in the second round. Otherwise, the consensus cannot be achieved even if $m>\lfloor{K/2}\rfloor$. 

\begin{figure}
	\centering
	\includegraphics[scale=0.17]{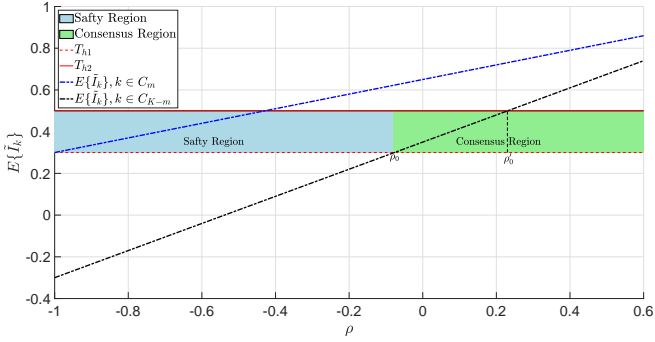}
	\caption{Consensus performance under fault tolerance, $\alpha=0.35$}
	\label{fig_cons_under_threshols}
\end{figure}

\begin{figure}
	\centering
	\includegraphics[scale=0.17]{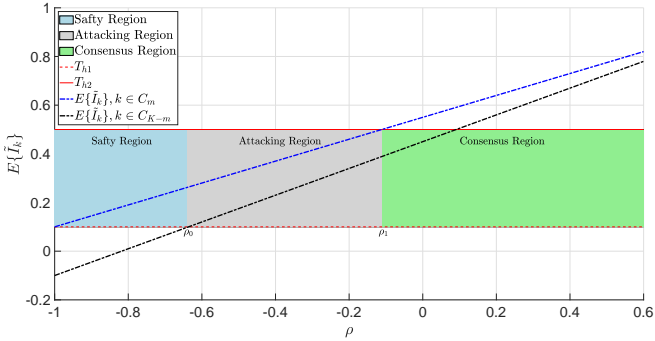}
	\caption{Consensus performance over fault tolerance, $\alpha=0.45$}
	\label{fig_cons_over_threshols}
\end{figure}

We show two examples in Fig.~\ref{fig_cons_under_threshols} and Fig.~\ref{fig_cons_over_threshols}. In Fig.~\ref{fig_cons_under_threshols}, we plot the HCF for all users with $\alpha=0.35$, which is under the threshold $\alpha^*$. The range of $\rho$ can be divided into two parts by the critical point $\rho_0$ that satisfies $\alpha+(1-\alpha)\rho_0=T_{h1}=1-2\alpha$. When $\rho<\rho_0$, the region is denoted as \emph{Safety Region}, all malicious users will be filtered by $T_{h1}$ and cannot enter the second round. When $\rho>\rho_0$, the region is denoted as \emph{Consensus Region}, malicious users can enter the second round, but they cannot affect the consensus results because the HCF of all honest users always exceeds $T_{h2}=0.5$ such that all honest users can enter \emph{Reply Phase} and send the reply message to the primary user. Note that in the second round, if $\rho_0<\rho<\rho'_0$, the HCF of malicious users is less than $T_{h2}$ such that they cannot send the reply message. If $\rho>\rho'_0$, malicious users also send a reply message. However, regardless of whether malicious users can send a reply message or not, it will not affect the consensus result because all honest users always send the reply message. Therefore, malicious users cannot affect consensus results whatever $\rho$ is set when $\alpha=0.35$. 

In  Fig.~\ref{fig_cons_over_threshols},  we plot the HCF of all users with $\alpha=0.45$, which is over the threshold $\alpha^*$. In this case, we can divide the range of $\rho$ into three parts by two critical point $\rho_0$ and $\rho_1$, where $\rho_0$ satisfies $\alpha+(1-\alpha)\rho_0=T_{h1}=1-2\alpha$ and $\rho_1$ satisfies $1-\alpha+\alpha\rho_1=T_{h2}=0.5$. Similarly, when $\rho<\rho_0$, the consensus can be achieved because malicious users cannot enter the second round. When $\rho>\rho_1$, the consensus also can be achieved because the HCF of honest users always exceeds $T_{h2}$. However, when $\rho_0<\rho<\rho_1$, we denote it as \emph{Attacking Region}, malicious users can enter the second round, and the HCF of all honest users does not exceed $T_{h2}$, so the consensus cannot be achieved and the attack is successful in this region. 


\subsection{Two-round consensus procedure}
Based on the hash consistency verification scheme discussed in the previous subsection, we propose a two-round consensus procedure as follows.  

In the first round (\emph{Prepare phase}), all users send their hash symbols to the BS via the AirComp technique. The BS feeds back the superposed hash symbol vector $\mathbf{t}$ to all users. Each user $k$ calculates the HCF $\tilde{I}_k$ according to  \eqref{eq13} and compares it with the threshold $T_{h1} = 0.22$. If $\tilde{I}_k > T_{h1}$, then the user changes to the prepared state and enters the second round of the consensus process. 

In the second round (\emph{Commit phase}), only users who are in the prepared state send the hash symbols to the BS via the AirComp technique. The BS feeds back the new superposed hash symbol vector $\mathbf{t}'$ to all users. Each prepared user $k$ calculates the HCF $\tilde{I}_k$ once again using $\mathbf{t}'$ and compares with the threshold $T_{h2} = 0.5$. Only those users with $\tilde{I}_k > T_{h2}$ enter the \emph{Reply Phase} and return the reply message to the BS via the AirComp technique. The superposed reply message is forwarded to the primary user by the BS. 

The final consensus result is determined by the primary user based on the superposed reply message from the BS. The primary user also can calculate the HCF in \eqref{eq13} between its reply message and the superposed message. If the HCF exceeds $T_h=0.5$, which suggests that more than half of the total users return the consistent reply message, then the consensus is achieved.

\section{AirCon implementation based on LTE system}\label{sec:aircomp}

In this section, we consider the implementation of AirCon protocol based on the open-source srsLTE platform, which provides the standard LTE protocols. We present solutions to some of the critical issues for practical AirComp implementation. Specifically, two problems are considered for AirComp implementation: 1) Synchronization problem; 2) Uplink channel estimation and feedback problem.  

\subsection{Synchronization}
The first challenging problem for AirComp implementation is to achieve strict timing/frequency synchronization across different users so that accurate signal superposition can be obtained at the BS. 

Fortunately, the LTE system has a complete set of synchronization mechanisms that can meet the synchronization requirement of AirComp. Specifically, in the downlink of the LTE system, the BS broadcasts synchronization channels (including PSS and SSS) periodically. If a user wants to connect to the BS, the first step is to search for these two channels to get timing and frequency synchronization. As long as the user is connected to the BS, it will keep tracking the timing/frequency synchronization via PSS/SSS. The user also compensates for the frequency offset of uplink channels according to the estimated value from the downlink channels. 

In the uplink, a ``timing advance" mechanism is adopted by all users for timing synchronization with the BS. Specifically, when the user connects to the BS, the BS measures the propagation delay and calculates a timing offset, named ``Timing Advance (TA)", which is fed back to the corresponding user. Whenever the user transmits data to the BS, it should transmit at the time with an advance of TA so that the signals from different users arrive at the BS without timing offset. 

In addition, the physical layer of the LTE system is based on the  OFDM technique, which adopts a cyclic prefix (CP) to cope with the multipath interference problem. Due to the cyclic property, the CP can also be used for channel synchronization. That is, as long as the timing offset is within a CP, the impact of the timing offset can be treated as a channel phase shift and compensated as a part of channel fading. 

In summary, the existing synchronization mechanisms in the LTE system are sufficient for the implementation of AirComp, so there is no need to design a dedicated AirComp synchronization scheme in an LTE system, which is part of the reason we choose the LTE system as the implementation platform of the AirCon protocol.

\subsection{Channel estimation and feedback}\label{Channel estimation_scheme}
The purpose of uplink channel estimation is to cope with the channel fading problem via pre-compensation on the user side. In practice,  there are two different ways to obtain uplink CSI. One way is that the BS estimates the uplink channel and sends feedback to the corresponding user, the other way is that a user infers the uplink CSI from the downlink signals by leveraging the channel reciprocity of the TDD channel.  In this work, we choose the first solution since it can be used for both FDD and TDD systems.  

\begin{figure}
	\centering
	\includegraphics[scale=0.33]{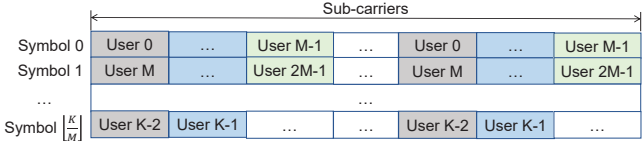}
	\caption{Assignment of reference symbols.}
	\label{fig_Reference_Signals}
\end{figure}

To get uplink CSI at the BS, a user needs to transmit reference symbols in the uplink channel. Ideally, the reference symbols should be transmitted at each sub-carrier so that the channel can be estimated accurately for all sub-carriers. However, this will consume too many channel resources and take a longer time for channel estimation for a large number of participating users. To this end, we exploit the property that the sub-carriers within coherence bandwidth may have a similar fading coefficient, so the reference symbols are only needed for every $M$ sub-carriers\footnote{$M$ is a parameter depending on the coherence bandwidth. Note that this scheme is also adopted by the uplink sounding reference signal (SRS) in the LTE system.}. As a result, at most $M$ users can share an OFDM symbol,  therefore a total of $\lceil{K/M}\rceil$ OFDM symbols are needed for reference symbol transmissions. Fig. \ref{fig_Reference_Signals} illustrates the assignment of the reference symbols for different users.  

Upon receiving the reference symbols, the BS extracts the received symbols and estimates the CSI in the order of user index one by one. The simplest estimation method is the least square (LS) method. Taking user 0 as a example, the BS firstly constructs the received symbol vector $\mathbf{y}=[y_0,y_M,\cdots,y_{\lfloor{N/M}\rfloor{M}}]$ from symbol 0, the corresponding transmitted reference signal for user 0 is $\mathbf{x}=[x_0,x_1,\cdots,x_{\lfloor{N/M}\rfloor}]$, then the LS estimation for uplink channel knowledge of user 0 can be expressed as
\begin{equation}\label{eq47}
	\tilde{\mathbf{h}}_{\text{LS}}=\left[\frac{y_0}{x_0},\frac{y_M}{x_1},\cdots,\frac{y_{\lfloor{N/M}\rfloor{M}}}{x_{\lfloor{N/M}\rfloor}}\right]^T
\end{equation}

However, the LS method does not consider the noise in the reference symbols. The LMMSE method\cite{Edfors_TC_98} can be employed to further reduce the noise impact by utilizing the cross-correlation between sub-carriers, which is given by \cite{Edfors_TC_98}:
\begin{equation}\label{eq46}
	\tilde{\mathbf{h}}_{\text{LMMSE}} = \mathbf{R}_{\mathbf{h}\mathbf{h}}(\mathbf{R}_{\mathbf{h}\mathbf{h}}+\frac{\beta}{\text{SNR}}\mathbf{I})^{-1}\tilde{\mathbf{h}}_{\text{LS}},
\end{equation}
where $\beta=\mathbb{E}\{|x_n|^2\}\mathbb{E}\{|1/x_n|^2\}$ is a constant depending on the reference symbol, 
$\mathbf{R}_{\mathbf{h}\mathbf{h}}=\mathbb{E}\{\mathbf{h}\mathbf{h}^H\}$ is the channel autocorrelation matrix\footnote{The matrix $\mathbf{R}_{\mathbf{h}\mathbf{h}}$ can be obtained from either a typical channel model\cite{Edfors_TC_98} or the channel LS estimation.}.

\begin{figure}[!t]
	\centering
	\includegraphics[scale=0.32]{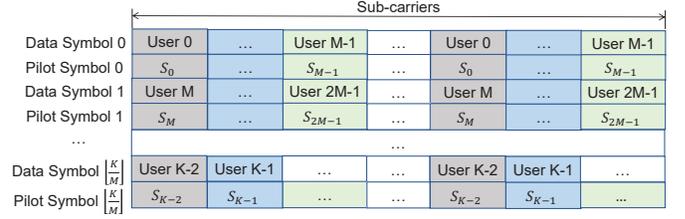}
	\caption{Assignment of downlink feedback symbols.}
	\label{fig_Feedback}
\end{figure}

Based on the channel estimation $\tilde{h}_{kn}$ for each subcarrier, the BS can set the coefficient of $\mathbf{B}_k$ as $b_{kn}={\tilde{h}^{\dagger}_{kn}}/{|\tilde{h}_{kn}|^2}$, which can compensate the channel fading and achieve the ideal signal aggregation as shown in Eq.\eqref{eq1} and Eq.\eqref{eq2}. This parameter should be fed back to the users as shown in Fig.~\ref{fig_Feedback}. Similar to the uplink channel estimation, two OFDM symbols are shared by every $M$ users, where the uplink CSI is transmitted in the \emph{Data Symbol} and downlink reference symbols are transmitted in the \emph{Pilot Symbol} for downlink channel estimation. Therefore, totally 2$\lceil{K/M}\rceil$ OFDM symbols are needed for downlink feedback transmissions.  We still take user 0 as example: the coefficient vector $\mathbf{b}_0=[b_{00},b_{0M},\cdots,b_{0~\lfloor{N/M}\rfloor{M}}]$ is transmitted at the corresponding sub-carrier in the \emph{Data Symbol} 0. In the \emph{Pilot Symbol} 0, the downlink reference symbol $s_0$ is transmitted at the corresponding sub-carrier. After receiving these two symbols, user $0$ firstly estimates the downlink channel using the received symbols from \emph{Pilot Symbol} 0. Then the user can obtain the coefficient vector $\mathbf{b}_0$ from \emph{Data Symbol} 0 by canceling the impact of downlink channel fading. Since each user only utilizes $1/M$ sub-carriers, the coefficients for the rest of sub-carriers can be estimated by interpolation\cite{Zhaoyp_VTC_97,Dongxd_TWC_07} at the users. 

In the low-SNR scenario, in addition to the channel estimation algorithm, a retransmission scheme (i.e., the uplink reference symbols and downlink feedback symbols are transmitted multiple times) can be adopted to further improve the performance. The impact of retransmission times on the consensus accuracy will be studied in section \ref{sec:performance}.

\subsection{AirCon implementation}

\begin{figure}[!t]
	\centering
	\includegraphics[scale=0.4]{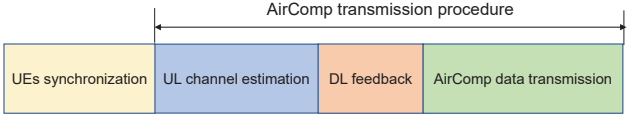}
	\caption{AirComp signaling and data transmission procedures.}
	\label{fig_process}
\end{figure}

The transmission procedure of the  AirComp protocol is summarized in Fig.~\ref{fig_process}, which consists of all procedures discussed in previous subsections. Firstly, the synchronization between the users and the BS will be established. Then the BS estimates uplink CSI for all users based on the uplink reference signals. The coefficients of  $\mathbf{B}_k$ are computed based on the channel estimation results and fed back to all users. Finally, all users transmit data using the AirComp technique. 

We implement the AirCon protocol based on the above AirComp procedures in the LTE system. The AirCon protocol is started when all users are in a stable state, such as the RRC IDLE state. In our implementation, we mainly focus on the feasibility of AirComp technology on hash verification of blockchain consensus, so only hash bits are generated in each user, rather than broadcasting an entire block from the primary user. In this case, we do not appoint a specific primary user. The consensus request is generated in the BS, which is a dedicated system information block (SIB) to notify all users that the BS is ready for consensus and users can transmit an uplink reference signal for AirComp uplink channel estimation. Since a user usually does not receive SIBs when it is in RRC IDLE state, the BS can set the \emph{systemInfoModification} field in the paging message (all users must receive the paging message periodically) to ensure all users receive the SIB (i.e., the consensus request). Similarly, the consensus result is determined by the BS rather than the primary user.

\subsection{Complexity analysis of the AirCon protocol}
\begin{figure}[!t]
	\centering
	\includegraphics[scale=0.4]{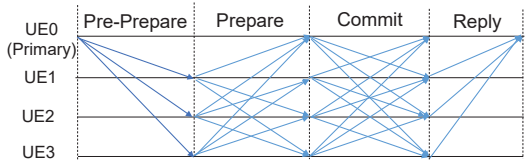}
	\caption{Traditional PBFT consensus diagram.}
	\label{fig_pbft}
\end{figure}

The complexity of the consensus protocol in wireless networks mainly comes from two aspects: communication complexity and computation complexity. As a benchmark, the traditional PBFT protocol is also illustrated in Fig.~\ref{fig_pbft} for the convenience of complexity comparison. We mainly focus on the complexity in the \emph{Prepare} and \emph{Commit} phases.
The total number of messages in the \emph{Prepare} and \emph{Commit} phases of the traditional PBFT protocol is 
\begin{equation}\label{eq42}  
	\text{Num}^{\text{message}}_{\text{PBFT}}=(K-1)(K-1)+K(K-1)=(2K-1)(K-1)
\end{equation}

Assuming each message consumes $N$ wireless resource blocks (RBs), then the number of wireless RBs consumed by the PBFT protocol is
\begin{equation}\label{eq43}  
	\text{Num}^{\text{RBs}}_{\text{PBFT}}=2N\cdot\text{Num}^{\text{message}}_{\text{pbft}}=2N(2K-1)(K-1)
\end{equation}
where the RE consumption in both the uplink and downlink directions is considered. 

For the AirCon protocol, the total number of wireless RBs consumed in the \emph{Prepare} and \emph{Commit} phases is
\begin{equation}\label{eq44}  
	\text{Num}^{\text{REs}}_{\text{AirCon}}=4N
\end{equation}

Therefore, the communication complexity of the traditional PBFT protocol is $\mathbb{O}(K^2)$,  while the communication complexity of the AirCon protocol is $\mathbb{O}(1)$.

If the overhead of channel estimation (CE) and feedback are also considered, the extra overhead is $4\lceil{K/M}\rceil N$, which has a complexity of $\mathbb{O}(K)$. In this case, the total number of consumed RBs is 
\begin{equation}\label{eq45}  
	\text{Num}^{\text{REs}}_{\text{AirCon\_CE}}=4N+4\lceil{K/M}\rceil N
\end{equation}

\begin{figure}[!t]
	\centering
	\includegraphics[scale=0.2]{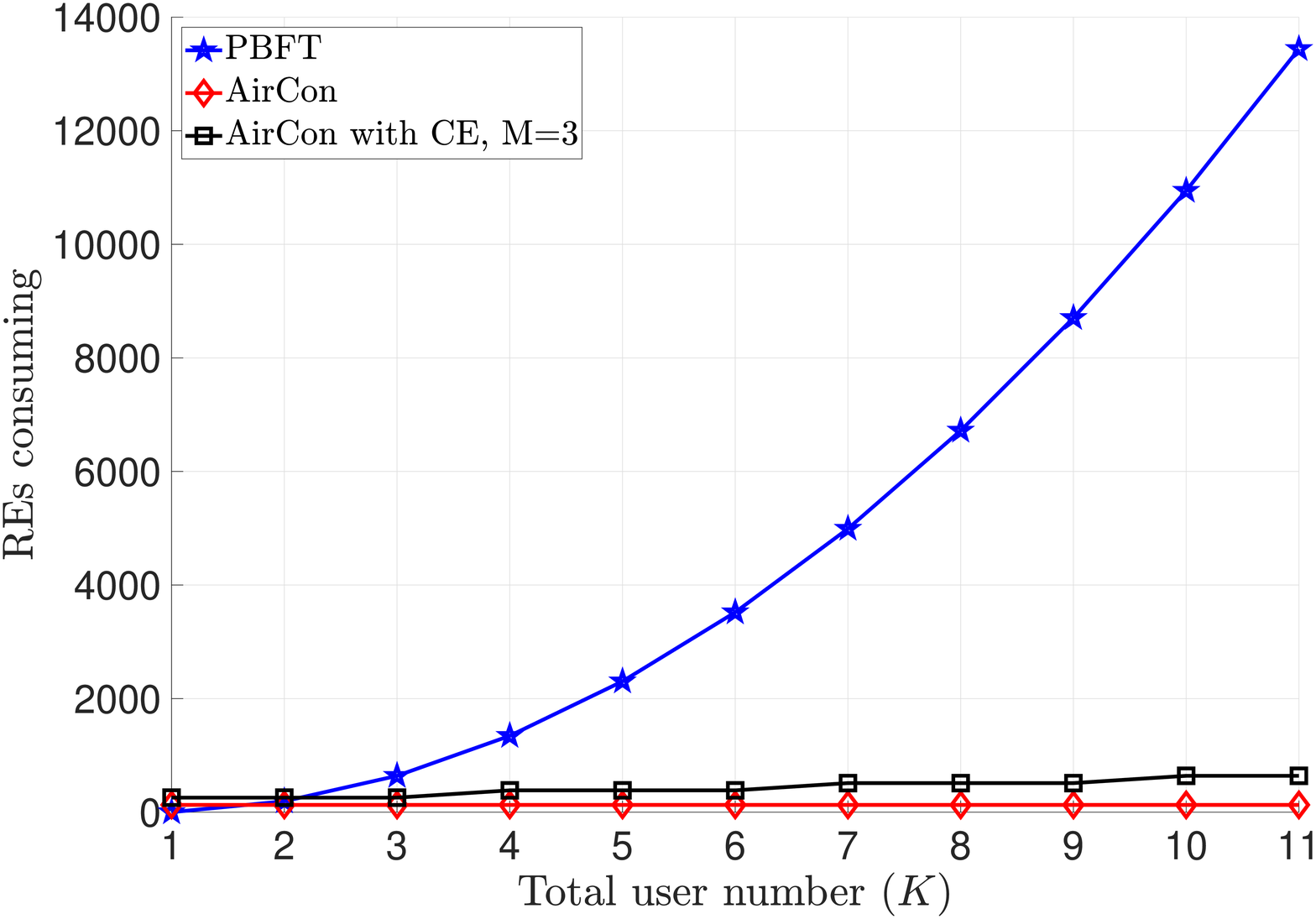}
	\caption{Consensus protocol REs consuming.}
	\label{fig_consensus_complexity}
\end{figure}

Note that the complexity of channel estimation depends on the system. If channel reciprocity is utilized to estimate the uplink channel, the complexity of channel estimation can be reduced to $\mathbb{O}(1)$. The RB consumption in both consensus protocols under different users number is illustrated in Fig.\ref{fig_consensus_complexity}.

As for the computational complexity, the hash verification procedures in the traditional consensus protocol are carried out in the application layer. Therefore, demodulation, decoding in the physical layer, and decryption in the upper layer are required. However, in our AirCon protocol, the consensus procedures are conducted in the physical layer, and the demodulation/decoding procedures are not required using the proposed hash consistency verification method. Therefore, the computational complexity only comes from $\mathbb{O}(K)$ multiplication operations. 

In summary, compared with the traditional PBFT protocol, the AirCon protocol has much lower communication overhead and computational complexity.

\section{Performance Evaluation}\label{sec:performance}
In this section, we show the implementation of the AirCon testbed and provide experimental results to demonstrate the performance under the real-world testbed. Due to the limitation of testbed conditions, we also present simulation results to evaluate the performance of the AirCon protocol under more general channel conditions.

\subsection{AirCon testbed design}
We implement the AirCon protocol based on srsLTE (version 20.04), an open-source LTE platform consisting of complete UE/eNodeB protocol stacks and a lightweight core network (CN) protocol stack. 
The software defines radio (SDR) board USRP B210\cite{USRP_B210} is used to transmit and receive RF signals. Based on the srsLTE platform, the following modules are re-used for AirComp implementation: (1) user attaching procedures, which guarantees all users are synchronized to the eNodeB; (2) The OFDM waveform generation, which can generate waveform based on the LTE frame structure. We then develop all other modules related to the AirCon protocol, which include: (1) reference symbols, hash bits generation, and lattice code modulation; (2) channel estimation module; (3) downlink feedback module; (4) hash consistency verification module and (5) AirCon protocol control module. 

\begin{figure}
	\centering
	\includegraphics[scale=0.30]{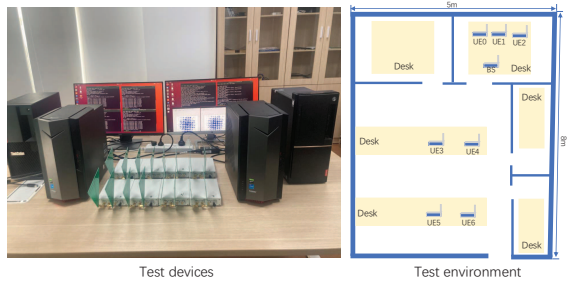}
	\caption{Testbed components and environment}
	\label{fig_test_bed}
\end{figure}

All devices of the testbed and setup of the test environment are shown in Fig.~\ref{fig_test_bed}. Totally eight USRP boards are used in our experiment, where one board is used as BS and the other seven boards are used as users. All these USRP boards are driven by four PCs via USB3.0 interfaces. 

For each experiment setup, the consensus procedures are repeated 1000 times. For a given number of $m$ users with consistent hash symbols, we define  consensus error ratio ($\text{CER}(m)$) as a measure of consensus performance, which is given by:
\begin{equation}\label{eq25}
	\text{CER}(m) = \frac{\text{Number of consensus errors}}{\text{Total number of consensuses}}
\end{equation}
whereby a consensus error corresponds to one of the following two cases: (1) The consensus is achieved when $m\le\lfloor{\frac{K}{2}}\rfloor$; (2) The consensus is not achieved when $m>\lfloor{\frac{K}{2}}\rfloor$.

\subsection{Testbed results}

\begin{figure}
	\centering
	\includegraphics[scale=0.2]{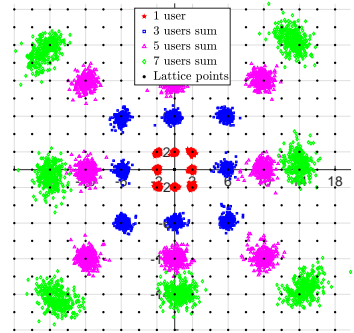}
	\caption{Symbol superposition when all users transmit same hash symbols, the received symbols are scaled with the number of users}
	\label{all_users_same}
\end{figure}

\begin{figure}
	\centering
	\includegraphics[scale=0.2]{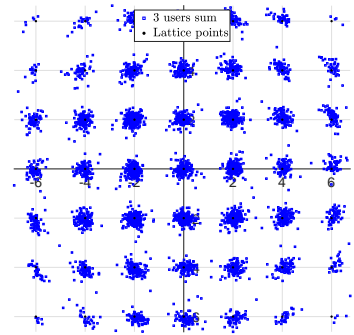}
	\caption{Symbol superposition when three users transmit different hash symbols, the received symbols fall on all fine lattice points}
	\label{3users_different}
\end{figure}

Firstly, we demonstrate the symbol superposition results of the AirComp technique. As shown in Fig.~\ref{all_users_same}, if all users transmit the same hash symbols, the received symbols at the BS are scaled with the number of users.  On the contrary, as shown in Fig.~\ref{3users_different}, if all users transmit different hash symbols, the superposed symbols should be distributed on all lattice nodes randomly. From these two figures, it can be seen that the channel fading is effectively eliminated using the proposed AirComp implementation schemes (in particular the channel estimation, feedback, and pre-compensation schemes), and the desired channel superposition is achieved. Therefore, it is feasible to apply the AirComp technology in a digital communication system.

\begin{figure*}
	\centering	
	\subfloat[\emph{Prepare} stage]{\includegraphics[scale=0.17]{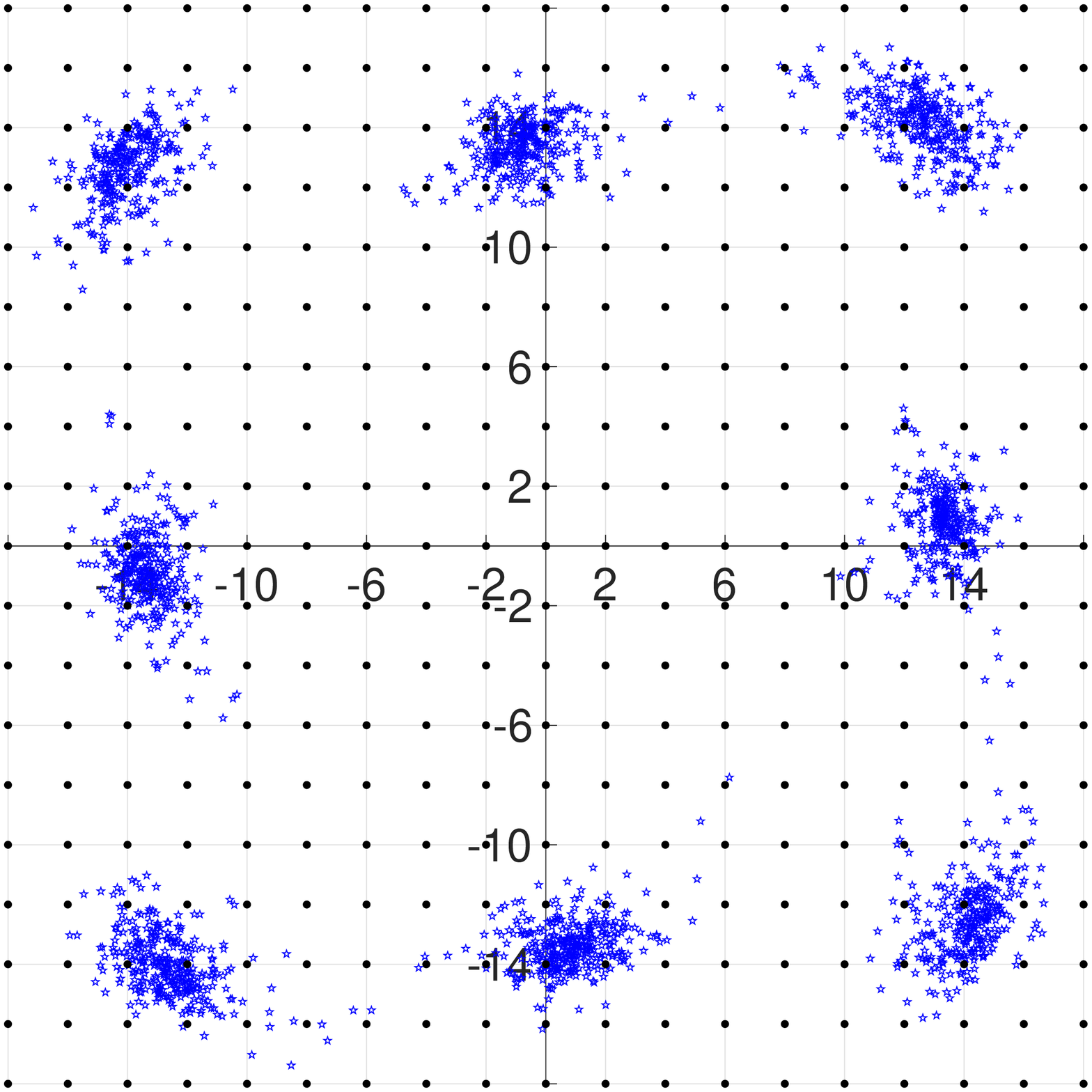}
		\label{7users_same1}}
	\hspace{1in}
	\subfloat[\emph{Commit} stage]{\includegraphics[scale=0.17]{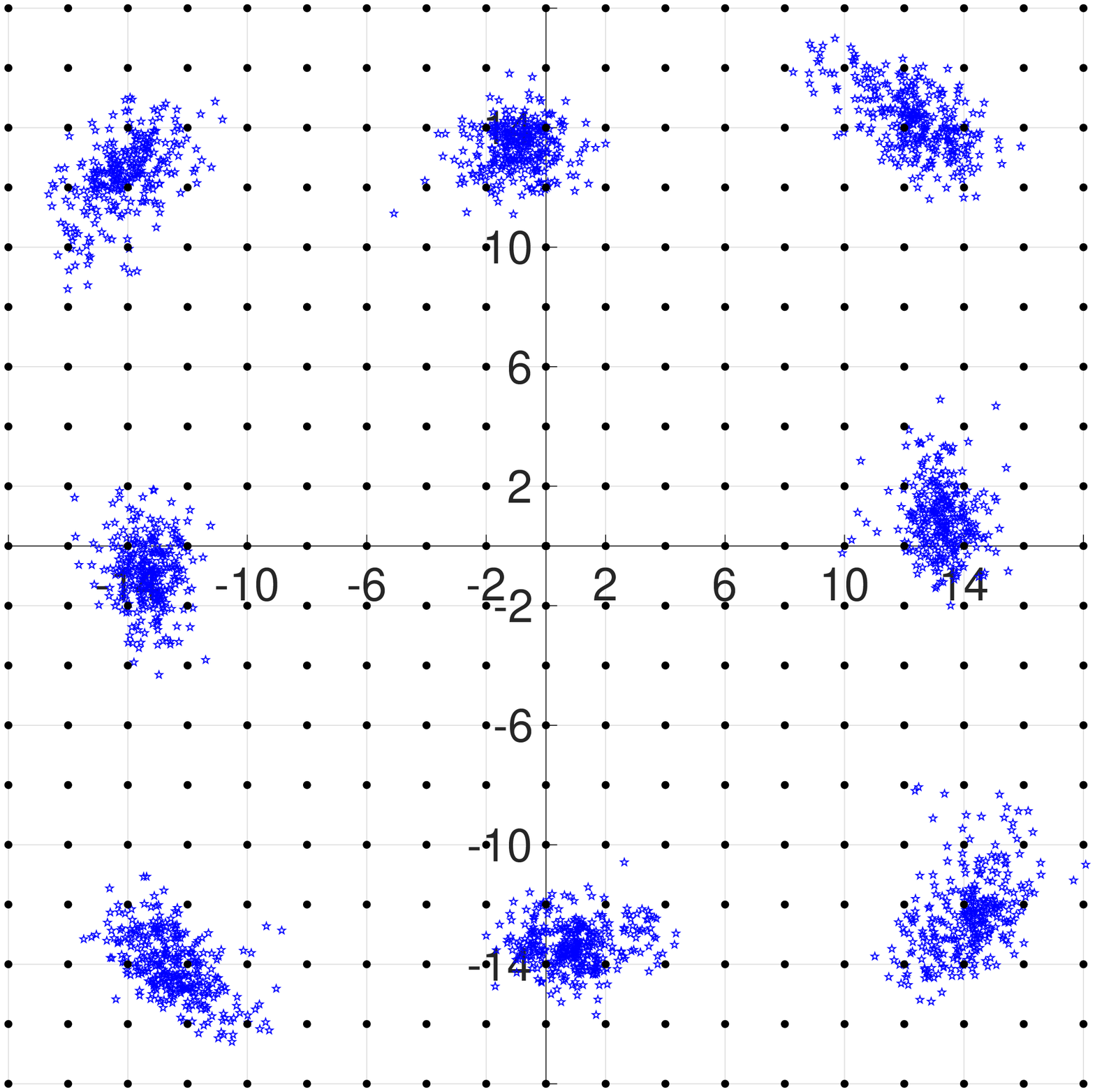}
		\label{7users_same2}}
	\hspace{1in}
	\subfloat[\emph{Reply} stage]{\includegraphics[scale=0.17]{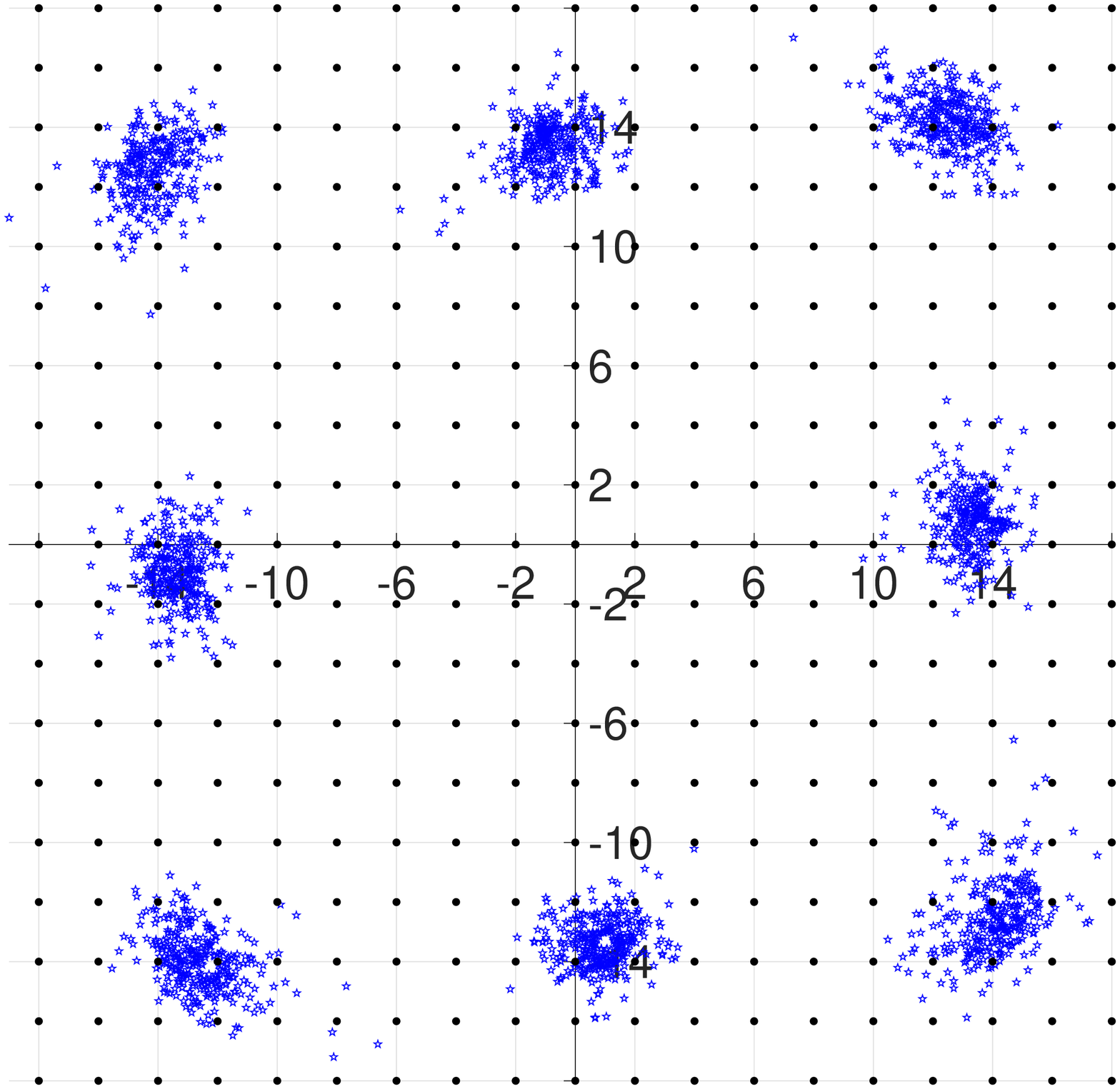}
		\label{7users_same3}}
	
	\caption{Seven users transmit consistent hash symbols, all users determine the final consensus result}
	\label{fig_testbed_results2}
\end{figure*}

\begin{figure*}
	\centering
	\subfloat[\emph{Prepare} stage]{\includegraphics[scale=0.17]{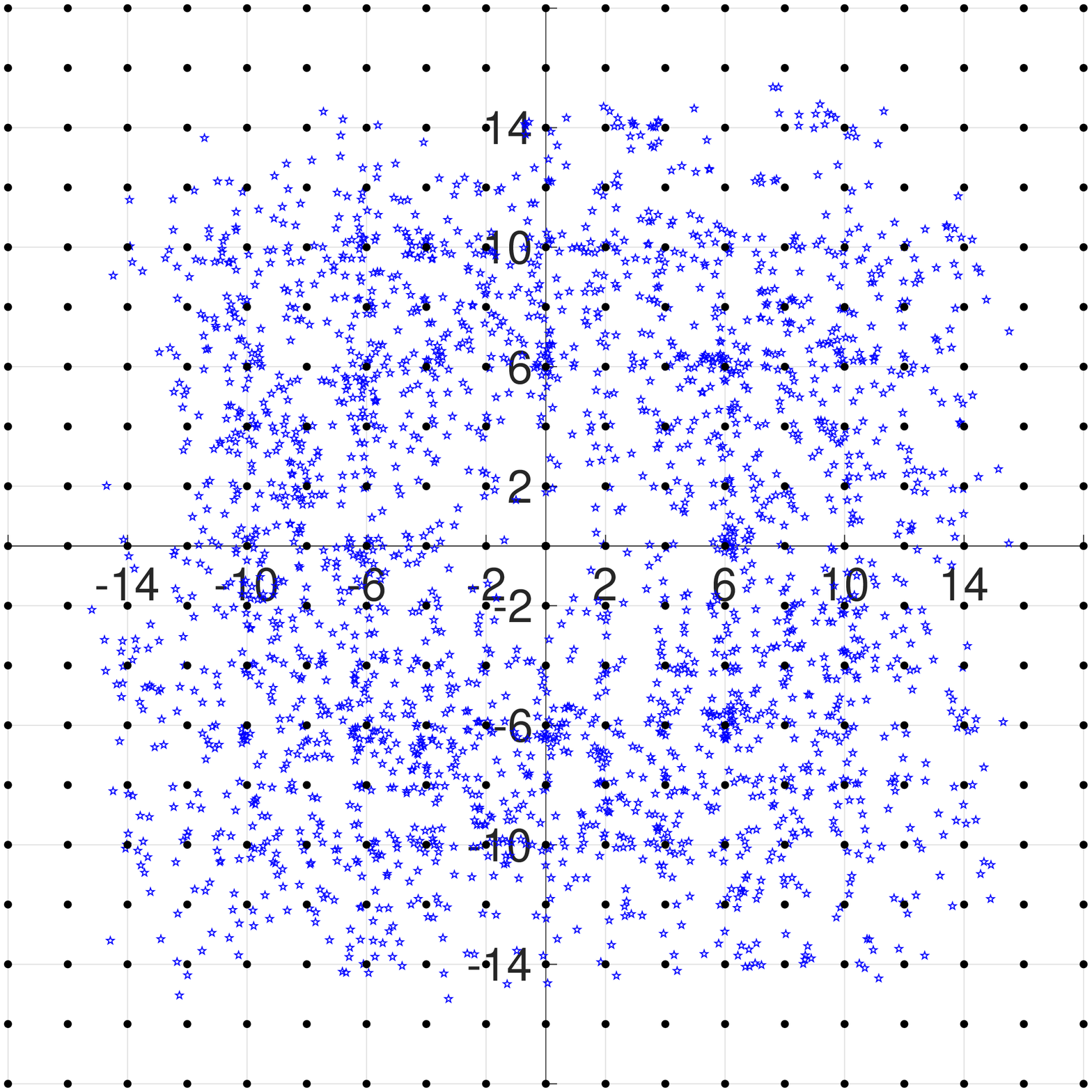}
		\label{4users_same1}}
	\hspace{1in}
	\subfloat[\emph{Commit} stage]{\includegraphics[scale=0.17]{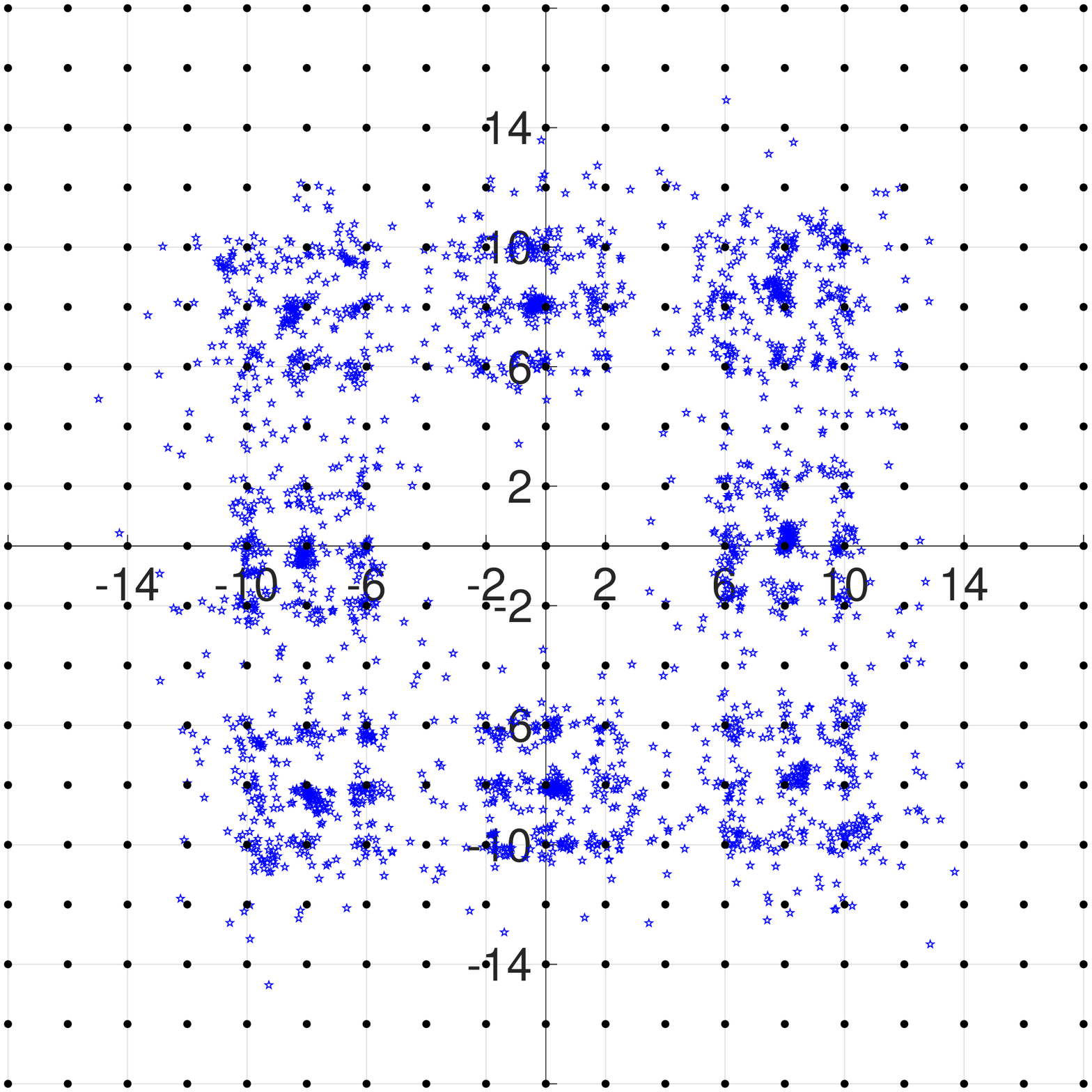}
		\label{4users_same2}}
	\hspace{1in}
	\subfloat[\emph{Reply} stage]{\includegraphics[scale=0.17]{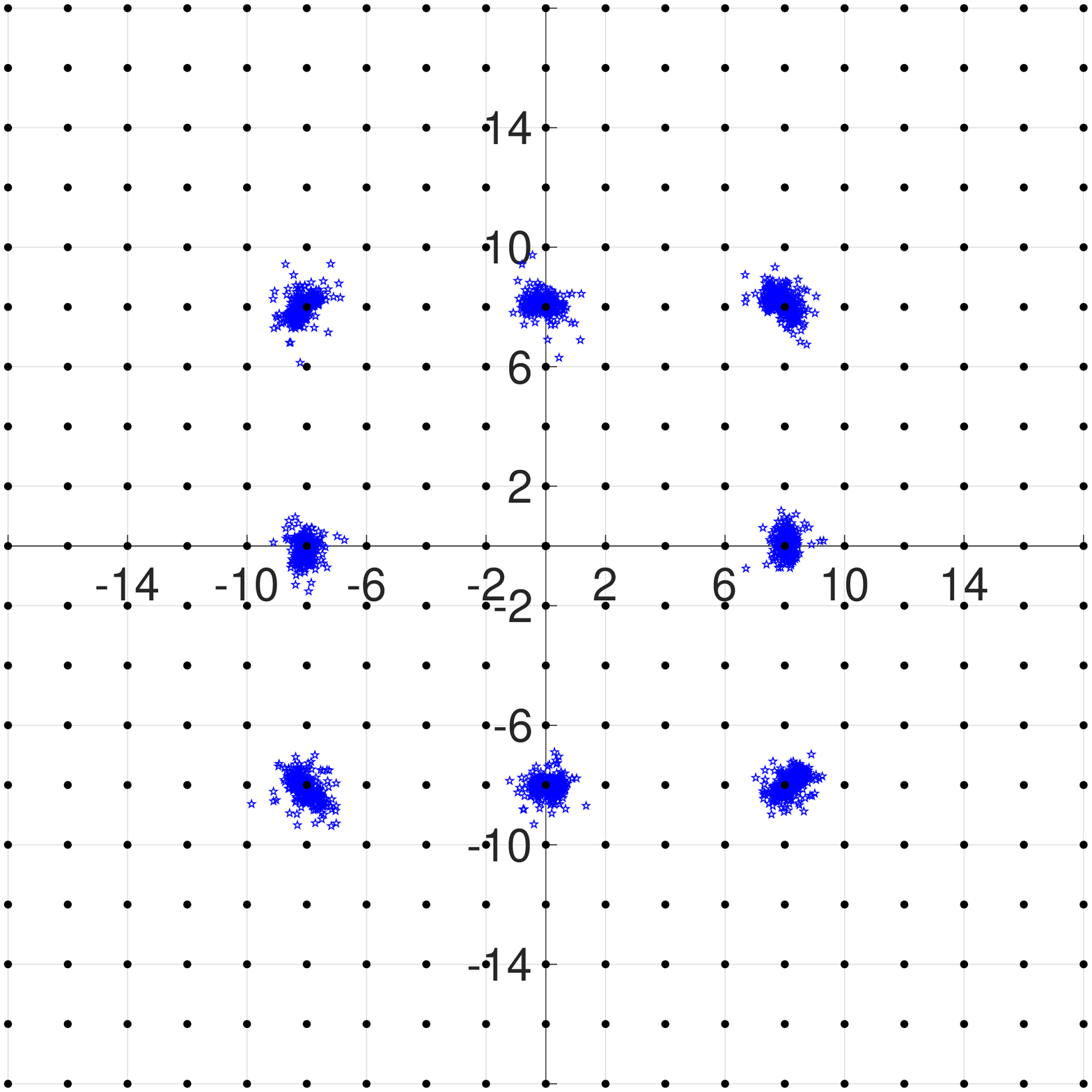}
		\label{4users_same3}}
	
	\caption{Four users transmit consistent hash symbols, three users transmit random hash symbols, those users with random hash symbols are filtered}
	\label{fig_testbed_results3}
\end{figure*}

Secondly, we show the superposed hash symbols at the BS at different consensus stages. In Fig.~\ref{fig_testbed_results2}, we show the case that all users transmit consistent hash symbols. It can be seen that all users can get into the {\it Commit} and {\it Reply} stages. In the second case, only four users transmit consistent hash symbols, while the other three users transmit random hash symbols. As shown in Fig.~\ref{fig_testbed_results3}, these three users with random hash symbols are filtered in the {\it Commit} stage, and only four users with consistent hash symbols enter the {\it Reply} stage.  

\begin{figure}
	\centering
	\includegraphics[scale=0.15]{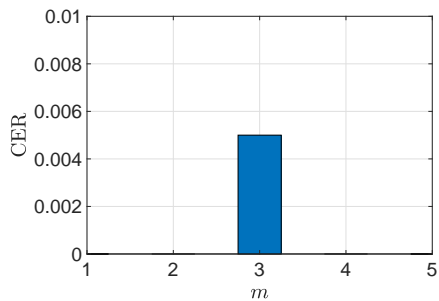}
	\caption{Consensus error ratio for real testbed result with $K=5$}
	\label{fig_testbed_cons_results}
\end{figure}

\begin{figure}
	\centering
	\includegraphics[scale=0.15]{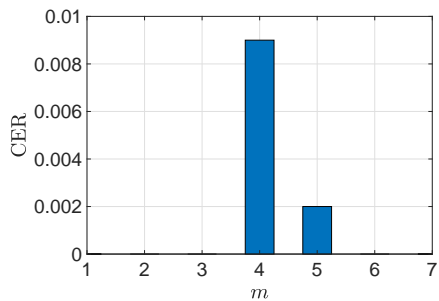}
	\caption{Consensus error ratio for real testbed result with $K=7$}
	\label{fig_testbed_cons_results2}
\end{figure}

The performance of the AirCon protocol are shown in Fig.~\ref{fig_testbed_cons_results} and Fig.~\ref{fig_testbed_cons_results2} with different users number ($K$). For $K=5$, UE5 and UE6 in Fig.~\ref{fig_test_bed} are excluded in the experiments. It can be observed that when the number of users with consistent hash symbols ($m$) is close to the threshold for reaching a consensus ($m=3$ for $K=5$ and $m=4$ for $K=7$), it may lead to consensus error. However, even in the worst case, the consensus error is lower than $1\%$, which demonstrates that the proposed AirCon protocol is reliable in practice. By carefully examining the experimental results, we observe that the consensus error is mainly caused by inaccurate downlink feedback.

\subsection{AirCon simulation setup}
We further evaluate the performance of the proposed AirCon protocol under more general conditions using simulations. In the simulation, the frame structure is the same as the practical LTE system with a 1.92MHz sampling rate. The channel gains for different users are generated independently.

In our simulation, three wireless channels are considered: 1) AWGN; 2) flat fading; 3) EPA (Extended Pedestrian A model), which is a multi-path channel model defined in the 3GPP TS 36.104\cite{3gpp_36_104} for the typical pedestrian wireless environments. The path delay and the corresponding delay power of the EPA channel are shown in Table~\ref{ch_tab}. 
We introduce the average CER (ACER) as a metric to compare the performance difference among different simulation setups, which is given by
\begin{equation}\label{eq26}
	\text{ACER} = \frac{1}{K}\sum_{m=1}^{K}\text{CER}(m)
\end{equation}

\begin{table}[]
	\centering
	\caption{Parameters of EPA channel}
	\label{ch_tab}
	\begin{tabular}{|l|l|l|l|l|l|l|l|}
		\hline
		Delay (ns) & 0 & 30 & 70 & 90 & 110 & 190   & 410   \\ \hline
		Power (dB) & 0 & -1 & -2 & -3 & -8  & -17.2 & -20.8 \\ \hline
	\end{tabular}
\end{table}

\subsection{Simulation results}\label{subsec:SimRslt}
\begin{figure}
	\centering
	\includegraphics[scale=0.17]{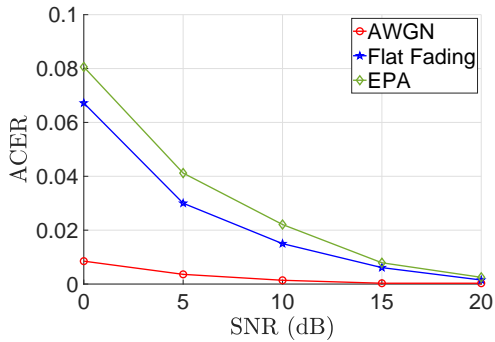}
	\caption{Consensus error ratio under different SNR ($K=11$)}
	\label{fig_SNR_Cons_Performance}
\end{figure}

In Fig.~\ref{fig_SNR_Cons_Performance}, we show the ACER of the AirCon protocol under different channel conditions. Firstly, it can be seen that the consensus performance is degraded among all types of channels when SNR decreases. Secondly, the performance gap between different types of channels is increasing in the low-SNR region, which suggests that the channel estimation accuracy loss is the dominant factor of the consensus performance loss in the low-SNR region. If the channel fading is well pre-compensated (e.g., an AWGN case), the average consensus error ratio of AirCon is about $1\%$ even if the SNR is only 0 dB.

\begin{figure}
	\centering
	\includegraphics[scale=0.17]{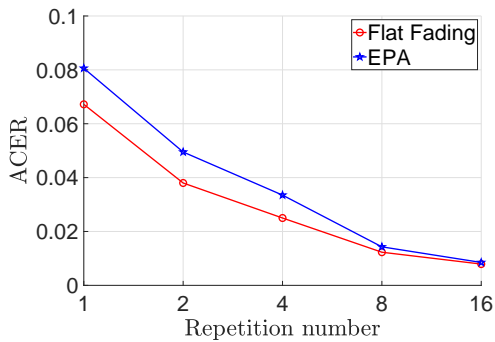}
	\caption{Consensus error ratio under different repetition number (SNR = 0 dB)}
	\label{fig_RepNum_Cons_Performance}
\end{figure}

In the low-SNR region, the retransmission scheme can help improve consensus accuracy. In Fig.~\ref{fig_RepNum_Cons_Performance}, we show the consensus performance improvement under different retransmission numbers. It can be observed that the consensus error ratio decreases gradually with the increase of the retransmission number. Note that the retransmission will consume more wireless resources. Therefore, the trade-off between retransmission number and consensus performance needs to be considered in practice. 

\begin{figure}
	\centering
	\includegraphics[scale=0.17]{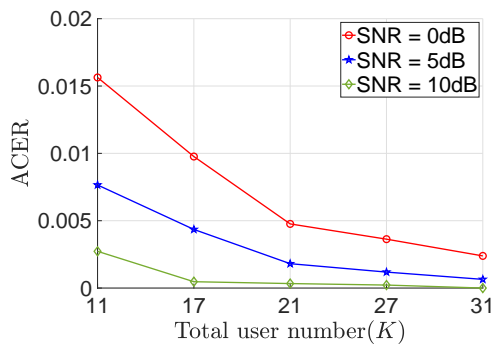}
	\caption{Averaging consensus error ratio under different $K$ (AWGN channel)}
	\label{fig_User_Num_Cons_Performance}
\end{figure}

\begin{figure}
	\centering
	\includegraphics[scale=0.17]{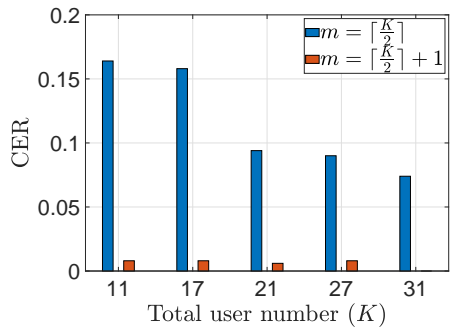}
	\caption{Consensus error ratio around threshold under different $K$ (SNR = 0 dB)}
	\label{fig_User_Num_Cons_Performance2}
\end{figure}

We also study the influence of total user number ($K$) on consensus accuracy. Due to space limitation, we only show the AWGN channel case. It can be observed from Fig.~\ref{fig_User_Num_Cons_Performance} that the AirCon protocol performs better when $K$ increases. The improvement of consensus performance comes from two aspects: (1) The consensus error only appears when $m$ is around the threshold for reaching the consensus (namely, $m=\lceil{\frac{K}{2}}\rceil$ and $m=\lceil{\frac{K}{2}}\rceil+1$ in our simulation). Therefore, the proportion of cases that will not incur consensus error increases when the total number of users increases. (2) Even in the cases that $m=\lceil{\frac{K}{2}}\rceil$ and $m=\lceil{\frac{K}{2}}\rceil+1$, the consensus error ratio also decreases, which is illustrated in Fig.~\ref{fig_User_Num_Cons_Performance2}. That is because when more users participate in the AirComp process, the average effect of noise on each user in the superposed signal is reduced, so the consensus performance becomes better.

\section{Conclusion}\label{sec:conclusion}
In this paper, we have proposed the AirCon to achieve low complexity consensus for blockchain-enabled wireless networks, which is novel in that the hash symbols of all users are transmitted to the  BS  simultaneously over the same wireless spectrum via the  AirComp and lattice coding techniques, and the consensus can be done in the physical layer without decoding the hash symbols. We have shown that the AirCon protocol can significantly reduce the transmission and computational complexity during the consensus procedure. We have shown the design of AirCon based on the universal LTE system and provided solutions for the critical issues involved in the AirComp implementation, such as channel estimation and feedback. We also have implemented AirCon based on a srsLTE testbed, and real-world measurement results demonstrate that the proposed AirCon protocol
is feasible in a practical LTE system. In addition, we have also conducted extensive simulation experiments and shown the accuracy and robustness of the proposed consensus protocol under different network conditions. The proposed AirCon is inspiring for designing efficient consensus schemes over wireless networks, where a more efficient channel status acquiring scheme can bring significant performance gain for AirCon. 

\ifCLASSOPTIONcompsoc
\else
\fi

\ifCLASSOPTIONcaptionsoff
  \newpage
\fi



%
\bibliographystyle{IEEEtran}
\bibliography{AirCon}


%








\end{document}